\documentclass[12pt]{amsart}
\usepackage{amsmath,amstext,amsfonts,amsthm,amssymb}
\usepackage{eucal}
\usepackage{graphicx}
\renewcommand{\subsubsection}[1]{\addtocounter{subsubsection}{1}
{\ \\[3pt]\bf \thesubsubsection. \  #1} }

\swapnumbers

{  \theoremstyle{definition}

}
\newcommand{\hra}{\hookrightarrow}

\newcommand{\lra}{\longrightarrow}
\newcommand{\ra}{\rightarrow}
\newcommand{\lla}{\longleftarrow}
\newcommand{\dpar}{\partial}

\newcommand{\bea}{\begin{eqnarray*}}
\newcommand{\eea}{\end{eqnarray*}}
\newcommand{\bean}{\begin{eqnarray}}
\newcommand{\eean}{\end{eqnarray}}

\newcommand{\bn}{{\bar{n}}}
\newcommand{\bV}{{\bar{V}}}

\newcommand{\tf}{\tilde f}

\newcommand{\tpsi}{\tilde\psi}

\newcommand{\fg}{\mathfrak g}

\newcommand{\fh}{\mathfrak h}

\newcommand{\fn}{\mathfrak n}

\newcommand{\fgl}{\mathfrak{gl}}
\newcommand{\fsl}{\mathfrak{sl}}

\newcommand{\CI}{\mathcal{I}}

\newcommand{\CL}{\mathcal{L}}
\newcommand{\CM}{\mathcal{M}}
\newcommand{\CO}{\mathcal{O}}
\newcommand{\CP}{\mathcal{P}}
\newcommand{\CQ}{\mathcal{Q}}

\newcommand{\CX}{\mathcal{X}}

\newcommand{\C}{\mathbb{C}}
\newcommand{\BC}{\mathbb{C}}

\newcommand{\BF}{\mathbb{F}}
\newcommand{\BN}{\mathbb{N}}
\newcommand{\BP}{\mathbb{P}}
\newcommand{\BQ}{\mathbb{Q}}

\newcommand{\BZ}{\mathbb{Z}}

\newcommand{\nc}{\newcommand}

\let\der\partial

\nc{\Id}{\text{Id}}
\nc{\la}{\lambda}

\begin{document}


\centerline{CONFORMAL BLOCKS AND EQUIVARIANT COHOMOLOGY}

\bigskip\bigskip


\vspace{1cm}

\centerline{Richard Rim\'anyi\footnote{Supported by the Marie Curie Fellowship PIEF-GA-2009-235437 and NSA grant CON:H98230-10-1-0171}, Vadim Schechtman and Alexander
Varchenko\footnote{Supported in part by NSF grant DMS-0555327}}

\vspace{2cm}

\hspace{6cm}{\it To the memory of Vladimir Arnold}

\vspace{2cm}

\bigskip\bigskip

\centerline{\bf \S 1. Introduction. On multinomial coefficients}

\bigskip\bigskip

{\bf 1.1.} 
Let $\bar V$ be the $m$-dimensional vector representation of the complex Lie algebra 
$\fg = \frak{gl}(m)$; consider its $N$-fold tensor product $V = \bar V^{\otimes N}$. The space $V$ is graded by the set 
$\CP_m(N)$ of  $m$-tuples of natural numbers 
$\lambda = (\lambda_1,\ldots,\lambda_m)$ with $\sum \lambda_i = N$: 
$$
V = \oplus_{\lambda\in\CP_m(N)}\ V_\lambda
\eqno{(1.1)}
$$
(for the definition of this gradation see $\S 2$ below; the reader may try to figure it 
out as an exercise). The dimension of $V_\lambda$ is given by the multinomial 
coefficient:
$$
\dim V_\lambda = C_\lambda := \frac{(\sum \lambda_i)!}{\lambda_1!\ldots\lambda_m!}.
\eqno{(1.2)}
$$
For example, if $m=2$, the decomposition (1.1) corresponds the familiar formula 
$(1+1)^N = \sum_{i=0}^N \binom{N}{i}$.

On the other hand the same numbers appear as the dimensions of certain cohomology. 
Namely, let $X_\lambda$ denote the variety of flags of linear subspaces 
$0 = L_0 \subset L_1 \subset \ldots \subset L_m = \BC^N$ where $\dim L_i/L_{i-1} = \lambda_i$; this is a smooth complex projective variety of dimension 
$$
d_\lambda = \sum_{i<j}\ \lambda_i\lambda_{j}.
$$
It has only even complex cohomology; consider the total cohomology 
space $H^*(X_\lambda) = \oplus_{i=0}^{d_\lambda} H^{2i}(X_\lambda)$ where 
by definition $H^k(X) := H^k(X,\BC)$. Then 
$$
\dim H^*(X_\lambda) = C_\lambda.
\eqno{(1.3)}
$$
To see (1.3) one can argue as follows, following Weil and Grothendieck. We consider $X_\lambda$ as the set of 
$\BC$-points of a $\BZ$-scheme $\mathcal X_{\lambda}$. Given a prime power $q = p^k$, the 
$\BF_q$-points of it are by definition flags in $\BF_q^N$, so their number 
is given by the $q$-multinomial coefficient
$$
\# \CX_{\lambda}(\BF_q) = C_\lambda(q) := \frac{[N]^!_q}
{\prod_{i=1}^m [\lambda_i]^!_q}
\eqno{(1.4)}
$$
where
$$
[n]_q^! = \prod_{i=1}^n\ [i]_q,\ [n]_q = \frac{q^n - 1}{q - 1}.
\eqno{(1.5)}
$$
Now we apply the Lefschetz fixed point formula in $\ell$-adic cohomology ($\ell\neq p$) 
to the $\bar\BF_q$-variety $X_{\lambda;q}:= \CX_\lambda\otimes_\BZ \bar\BF_q$: 
$$
\# \CX_{\lambda}(\BF_q) = \sum_{i=0}^{d_\lambda} Tr(F_q;H^{2i}(\CX_{\lambda;q},\BQ_\ell)) = 
\sum_{i=0}^{d_\lambda} \dim H^{2i}(X_\lambda)q^i
\eqno{(1.6)}
$$
where $F_q$ is the Frobenius endomorphism; in our case it acts on 
$H^{2i}(\CX_{\lambda;q},\BQ_\ell)$ as the multiplication by $q^i$; we also use   
the comparison theorem of complex and $\ell$-adic  cohomology. This implies  
that the limit of (1.4) when $q \ra 1$ gives the left hand side of (1.3). 
(In the case of grassmanians ($m = 2$) the above argument is contained in  
Andr\'e Weil's classical paper [W]; of course the historical logic is opposite...) 

{\bf 1.2.} Instead of usual cohomology one can consider the equivariant one, and use 
another incarnation of the Lefschetz formula --- the Atiyah - Bott localization  
theorem. Namely, the complex torus $T = \BC^{*N}$ acts 
naturally on $X_\lambda$ with $C_\lambda$ fixed points. The $T$-equivariant cohomology 
$H^*_T(X_\lambda)$ is a commutative $R:= H^*_T(*)$-algebra where the last ring may be identified 
with the polynomial algebra
$$
R = \BC[z_1,\ldots, z_N],\ z_i = c_1(M_i),
\eqno{(1.7)}
$$
$M_i = \BC$ with $T$ acting through the $i$-th projection $T \lra \BC^*$. As before, 
all cohomology is even. One can show $H^*_T(X_\lambda)$ is a free $R$-module of rank 
$\dim H^*(X_\lambda)$.  
The Atiyah - Bott theorem gives a basis of this module after certain localization.

Namely, 
consider the localized algebra
$$
R' = R[D^{-1}],\ D = \prod_{i<j}\ (z_i - z_j). 
\eqno{(1.8)}
$$
Let $i_\lambda:\ X_\lambda^T \hra X_\lambda$ be the inclusion of the set of fixed points. 
The Atiyah - Bott localization theorem [AB] says that the restriction map
$$
i_\lambda^*:\ H^*_T(X_\lambda)_{R'} \lra H^*_T(X_\lambda^T)_{R'}   
\eqno{(1.9)}
$$
is an isomorphism\footnote{for an $R$-module $M$, $M_{R'} := M\otimes_R R'$}. We have  
$\# X_\lambda^T = C_\lambda$, whence in particular (1.3). One can say that in the first proof of (1.3) the dimensions have been deformed, whereas in the
second proof the vector spaces are deformed. 

In fact, we get more. One can define a bijection of $X_\lambda^T$ with a certain basis in $V_\lambda$, 
so we get an isomorphism of two free $R'$-modules of rank $C_\lambda$ 
$$
\phi_\lambda:\ V_{\lambda;R'} \overset{\sim}\lra H^*_T(X_\lambda)_{R'}
\eqno{(1.10)}  
$$
by identifying their respective bases. Using these bases one defines a canonical 
element 
$$
y_\lambda\in V_\lambda\otimes H^*_T(X_\lambda)_{R'}
$$
which we may integrate along $X_\lambda$ to obtain an element 
$p_\lambda\in V_\lambda\otimes R'$ which we may interprete as a rational $V_\lambda$-valued function in $z_i$'s.  

The first observation of the present note (see \S 2) is that $p_\lambda$ coincides with the 
element constructed in [RV] and thus satisfies all the nice properties of the last element. 
In particular if $\lambda$ is such that the corresponding "bundle of conformal blocks" 
is of rank $1$, for example   
$\lambda = (a,a,\ldots,a)$, then $p_\lambda$ satisfies the Knizhnik - Zamolodchikov differential equations (this is not true for a general $\lambda$). 
It seems also that the element $y_\lambda$ before the integration has 
some remarkable properties.  

One possible advantage of this construction is that it works for any other cohomology 
theory satisfying Atiyah - Bott: for example one can replace the usual cohomology 
by $K$-theory; in this case one should obtain a "$q$-difference" version of the picture. 

{\bf 1.3.} Secondly, we deal with the situation of rank $1$ conformal blocks. In that case 
we have two natural generating sections of this bundle: the first one coming from the 
equivariant cohomology and the second one given 
by a hypergeometric integral from [SV]. These two sections are proportional; 
the proportionality coefficient ("normalisation constant") is given as usual 
by a "period": a Selberg type 
integral, we compute these integrals in \S 4. These two ways to define 
conformal blocks are somewhat similar to two ways of defining the Givental 
hypergeometric functions connected with quantum cohomology of flag spaces: 
the first one via the integration of a certain canonical 
element in the cohomology of a quasimaps' space (cf. [G1, Br]), the second 
--- mirror dual --- one, via stationary phase integrals, cf. [G2]. 
This analogy with mirror symmetry was the starting point of our reflections.    

Finally in the last Section, \S 5, we define geometrically an action of the Lie algebra of 
positive currents $\fgl(m)[t]$ on the equivariant cohomology $H^*_T(X_{m,N})$ where
$X_{m,N} = \coprod_{\lambda\in\CP_m(N)}\ X_\lambda$ in such a way that the isomorphisms 
(1.10), summed over all $\lambda$, become $\fgl(m)[t]$-equivariant. 
This action seems to be closely related to the actions studied by Ginzburg, Nakajima and others, cf. 5.4. 

We are grateful to M.Finkelberg and V.Ginzburg for very useful consultations. 
This paper was written while the third author was visiting the Institut de 
math\'ematiques de Toulouse. He thanks this Institute for the hospitality.

\bigskip\bigskip



\centerline{\bf \S 2. Weight spaces and fixed points}

\bigskip\bigskip

{\bf 2.1.} {\it The gradation in $V$.} We identify the fundamental representation $\bar V$ of $\fg = \frak{gl}(m)$ with the component 
of degree $1$ in the polynomial algebra $\BC[y_1,\ldots,y_m]$ where $\deg y_i = 1$, 
with the obvious action of $\fg$. 

More generally, 
the $\fg$-module $V = \bar V^{\otimes N}$ will be identified with a subspace   
in the polynomial ring in $mN$ variables $y_{ij}$, $1\leq i \leq m,\ 1\leq j\leq N$ spanned 
by all monomials 
$$
y^A = \prod_{i,j}\ y_{ij}^{a_{ij}}
$$
which for each $j = 1, \ldots, N$ contain exactly one character $y_{ij}$. In other words, 
the basis $\{ y^A\}$ is in one-to-one correspondence with the set $M(m,N)$ of $m\times N$ matrices 
$A = (a_{ij})$ whose entries are zeros or ones, which contain exactly one $1$ in each column.  

Given such a matrix $A$, we can do otherwise, and count the number of $1$'s in its rows. 
Set $\lambda(A) = (\lambda_1(A),\ldots,\lambda_m(A))$ where 
$$
\lambda_i(A) = \# \{j|\ a_{ij} = 1\}.
$$
Obviously $\lambda(A)\in \CP_m(N)$; we set $M(\lambda) := \{A\in M(m,N)|\ \lambda(A) = \lambda)$. In each set $M(\lambda)$ we define a point $M_\lambda\in M(\lambda)$ to be the 
matrix with
$$
(M_\lambda)_{ij} = 1\ \text{if\ } \mu_{i-1} \leq j\leq \mu_i
$$
where $\mu_i := \sum_{k=1}^i \lambda_k$.  

For example, for $\lambda = (1,1,\ldots,1)$, $M(\lambda) = $ 
the set of permutation matrices, $M_\lambda = $ the unity matrix.  

The symmetric group in $N$ letters $S_N$ acts on $M(\lambda)$ by permutation of columns; 
this action is transitive and 
the stabiliser of $M_\lambda$ is the subgroup
\newline  
$S_\lambda := S_{\lambda_1}\times\ldots\times S_{\lambda_m}$, which gives rise to 
a bijection 
$$
M(\lambda) \cong S(\lambda) := S_N/S_{\lambda}.
\eqno{(2.1)}
$$
It follows that $\# M(\lambda) = C_\lambda$. 

Another useful set in bijection with $M(\lambda)$ is defined as follows. 
Denote $[N] = \{1,2,\ldots,N\}$. Define $\Pi(\lambda)$ as the set of 
all maps $\pi:\ [N]\lra [m]$ such that $\#\pi^{-1}(i) = \lambda_i$ for all $i$. 
Given a matrix $M = (m_{ij})\in M(\lambda)$ let us associate to it a map $\pi$ as follows: 
we set $\pi(j) = i$ such that $m_{ij} = 1$; obviously $\pi\in\Pi(\lambda)$ and we've got a bijection
$$
M(\lambda)\cong\Pi(\lambda).
\eqno{(2.1a)}
$$   

Given $\lambda\in \CP_m(N)$, we define $V_\lambda\subset W$ to be the subspace 
spanned by the monomials $y^A$ with $A\in M(\lambda)$. 

{\bf 2.2.} {\it Cohomology of flag varieties.} Let $\lambda\in\CP_m(N)$. Recall the flag variety $X_\lambda$ of dimension $d_\lambda$. Over it we have the tautological flag 
of vector bundles 
$$
0 = \CL_0\subset \CL_1\subset\ldots\subset \CL_{m-1}\subset \CL_m = \CO_{X_\lambda}^N.
$$
Set $\CM_i := \CL_i/\CL_{i-1}$; these are vector bundles of dimension $\lambda_i$. 

The cohomology ring $H^*(X_\lambda)$ is generated as a graded $\BC$-algebra by the Chern  classes 
$c_{ij} := c_j(\CM_i)\in H^{2j}(X_\lambda),\ 1\leq i\leq m,\ 1\leq j\leq \lambda_i$, the ideal of relations is generated by $N$ relations which follow from the identity
$$
\prod_{i=1}^m\ (1 + \sum_{j=1}^{\lambda_j}\ c_{ij}t^j) = 1
\eqno{(2.2)}
$$
(i.e. we equate to $0$ all the coefficents of the $t$-polynomial on left, except the zeroth one). 

More generally, the $T$-equivariant cohomology may be described exactly in the same manner. Recall the coefficient ring $R = H^*_T(pt) = \BC[z_1,\ldots,z_N]$. As a graded $R$-algebra $H^*_T(X_\lambda)$ is generated by  the Chern classes 
$c_{T;ij} := c_j(\CM_i)\in H^{2j}_T(X_\lambda),\ 1\leq i\leq m,\ 1\leq j\leq \lambda_i$, the ideal of relations is generated by $N$ relations which follow from the identity
$$
\prod_{i=1}^m\ (1 + \sum_{j=1}^{\lambda_i}\ c_{T;ij}t^j) = 
\prod_{n=1}^N\ (1 + z_nt).
\eqno{(2.3)}
$$

It follows from this description that $H^*_T(X_\lambda)$ is a free graded $R$-module 
of rank $\dim H^*(X_\lambda)$. 

{\bf 2.3.} {\it Fixed points.} The action of $T$ on $X_\lambda$ has a finite set 
of fixed points $X^T_\lambda\subset X_\lambda$. To describe them explicitly, let 
$\{e_1,\ldots,e_N\}$ be the standard basis in $\BC^N$. Let 
$F_e = (0\subset F_1\subset\ldots\subset F_{m-1}\subset F_m = \BC^N)\in X_\lambda$ 
be the flag with $F_i$ being the subspace spanned by $e_1,\ldots, e_{\mu_i}$ 
(recall that $\mu_i = \sum_{j=1}^i\ \lambda_i$). Then $F_e$ is fixed 
under the action of $T$. 

The symmetric group $S_N$ acts on $\BC^N$ by permuting the elements of the standard basis, 
so it acts on the set of all flags. For all $\sigma\in S_N$ $F_\sigma := \sigma(F_e)$ 
belongs to $X_\lambda^T$ and in such a way we get all fixed points. The stabiliser of 
$F_e$ coincides with $S_\lambda$, so the mapping $\sigma\mapsto F_\sigma$ induces   
a bijection 
$$
S(\lambda) \cong X^T_\lambda, 
\eqno{(2.4)}
$$
cf. (2.1) 

For $w\in S(\lambda)$ we denote by $x_w$ the corresponding fixed point. The tangent 
space $T_w:=T_{X_\lambda,x_w}$ inherits the $T$-action; we will be interested in its Euler 
(top Chern) class: 
$$
e_w := c_{d_\lambda}(T_w)  \in H^{2d_\lambda}_T(pt).
$$
The explicit formula is as follows. Let $\pi_w\in\Pi(\lambda),\ \pi_w:\ [N]\lra [m]$, be the map corresponding 
to $w$ (cf. (2.1a)). Then 
$$
e_w = \prod_{i>j}\ \prod_{a\in\pi_w^{-1}(i), b\in\pi_w^{-1}(j)}\ (z_a - z_b).
\eqno{(2.5)}
$$

\bigskip

Let $i_w$ denote the inclusion $i_w:\ x_w\hra X_\lambda$; it is compatible with the 
$T$-action.
 
Define the elements $y'_{w} := i_{w*}(1)\in H^{2d_\lambda}(X_\lambda)$. The explicit formula for them is as follows. For each $i\in [m]$ let $\gamma_{ij}, j\in [\lambda_i]$, denote the {\it Chern roots} of $\CM_i$ --- the formal symbols such that 
$c_j(\CM_i) = \sigma_j(\gamma_{i1},\ldots,\gamma_{i\lambda_i})$, the elementary 
symmetric function. Let $\pi_w$ be as above. Then
$$
y'_w = \prod_{i > j}\prod_{a=1}^{\lambda_i}\prod_{b\in\pi_w^{-1}(j)}\ 
(\gamma_{ia} - z_b).
\eqno{(2.6)}
$$
Here is the main property of these elements which charterizes them: 
$$
i_w^*y'_{w'} = e_w\delta_{ww'}.
\eqno{(2.7)}
$$
The restriction map $i_w^*$ acts as follows: if $\pi_w^{-1}(i) = \{k_1,\ldots,k_{\lambda_i}\}$
with $k_1 < \ldots < k_{\lambda_i}$ then 
$$
i_w^*(\gamma_{ij}) = z_{k_j}.
$$
The composition $i_w^*i_{w*}$ equals the multiplication 
by $e_w$. 

Recall the localized ring $R'$; all $e_w$ become invertible in $R'$. 
The Atiyah - Bott localization theorem says that the restriction map is an isomorphism: 
$$
i^*:\ H^*_T(X_\lambda)':= H^*_T(X_\lambda)_{R'} \lra H^*_T(X^T_\lambda)' = 
\oplus_{w\in S(\lambda)}\ R'\cdot 1_w.
$$
The elements $y_w := y'_w/e_w$ form a basis of the free $R'$ module $H^*_T(X_\lambda)'$, 
cf. [AB] (the case $X_{(1,\ldots,1)} = G/B$ is discussed in [S]). 

Note that the explicit 
formula for $y_w$ written using (2.5) and (2.6) is very similar to the 
master function of a hypergeometric integral connected with a KZ equation. 

We shall also need the map $\int_{X_\lambda}:\ H^*_T(X_\lambda) \lra H^*_T(pt)$; 
we have for it $\int_{X_\lambda} y'_w = 1$.

\bigskip

{\bf 2.3.1.}
{\it Example.} For $X_\lambda = \BP^{m-1}$ (i.e. $\lambda = (m-1,1)$), 
$H^*_T(\BP^{m-1}) = \BC[x,z_1,\ldots,z_m]/(\prod (x - z_i))$ where $x = c_1(\CO(1))$. 
The action of $T$ has $m$ fixed points $x_i,\ i = 1, \ldots, m$; we have 
$y'_i = \prod_{j\neq i}\ (x - z_j)$, $e_i = \prod_{j\neq i}\ (z_i - z_j)$, 
$y_i = y'_i/e_i$ is nothing else but the $i$-th Lagrange interpolation polynomial.

\bigskip

{\bf 2.4.} {\it The canonical element.} Let us identify $M(\lambda)$ with 
$S(\lambda)$ by means of the bijection defined above. So for each $w\in S(\lambda)$ 
we will have the corresponding element in the weight subspace $y^w\in V_\lambda$ from 2.1 
on the one hand, and the element $y_w\in H^*_T(X_\lambda)'$ on the other hand. 

Consider the sum
$$
y_\lambda = \sum_{w\in S(\lambda)}\ y^w\otimes y_w\in V_\lambda\otimes H^*_T(X_\lambda)'.
$$
After integration over $X_\lambda$ we get an element 
$$
p_\lambda  := \int_{X_\lambda} y_\lambda = \sum_{w\in S(\lambda)}\  
\frac{y^w}{e_w} \in V_\lambda\otimes R' = V_\lambda[z_1,\ldots,z_N][D^{-1}].
$$
We may consider $p_\lambda = p_\lambda(z)$ as a rational function in variables $z_1,\ldots,z_N$ 
with poles along the diagonals, taking values in $V_\lambda$. 

To compare this element with that from [RV], let us recall some notation from 
there. Let $\CI$ denote the set of all decompositions of the set $[N]$ into 
a disjoint union  
$$
[N] = \coprod_{j=1}^m\ I_j
$$
with $\sharp I_j = \lambda_j$. We set
$$
R(z_{I_1}|z_{I_2}|\ldots |z_{I_m}) = \prod_{i<j} \prod_{a\in I_i, b\in I_j}\ 
(z_b - z_a). 
$$
Define 
$$ 
P_z(\lambda) = \sum_\CI\ \frac{\prod_{j=1}^m \prod_{a\in I_j}\ y_{ia}}
{R(z_{I_1}|z_{I_2}|\ldots |z_{I_m})},
$$
cf. [RV], Defintion 4.1 (up to a sign). 

{\bf 2.5.} {\it Theorem.} {\it The element $p_\lambda(z)$ coincides with the element $P_z(\lambda)$.} 

This is evident after identifying the Euler classes $e_w$ with the 
elements 
\newline $R(z_{I_1}|\ldots |z_{I_m})$. 

Therefore, $p_\lambda(z)$ satisfies all properties proven in {\it op. cit.} 
Let us list these properties. Let $\{ e_{ij},\ 1\leq i,j \leq m\}$ be the standard basis 
of $\fg$. For $x\in\fg$ we shall denote by $x^{(i)}$ the operator on $\bar V^{\otimes N}$ 
acting as $x$ on the $i$-th factor and identity on other factors. 

We define the {\it subspace of singular vectors} 
$$
V^s =  
\{ y\in V|\ e_{ij}y = 0,\ 1\leq i < j \leq m\}.
$$
Let us call $\lambda = (\lambda_1,\ldots,\lambda_m)\in \CP_m(N)$ {\it a partition} 
if $\lambda_1\geq \ldots\geq \lambda_m$; we denote by $\CQ_m(N) \subset \CP_m(N)$ 
the subset of all partitions. Denote $V_\lambda^s = V^s\cap V_\lambda$. We have 
$$
V^s = \oplus_{\lambda\in\CQ_m(N)}\ V_\lambda^s.
$$
We denote $Z = \{ (z_1,\ldots,z_N)\in \BC^N|\ z_i\neq z_j\ \text{for\ }i\neq j\}$. 
For $z = (z_1,\ldots,z_N)\in Z$ we denote by $e(z)$ the operator 
$$
e(z) = \sum_{i=1}^N\ z_ie_{1m}^{(i)}
$$
acting on $V$. Given a partition $\lambda = (\lambda_1,\ldots,\lambda_m)$ and 
a natural $\ell\geq\lambda_1 - \lambda_m$ ({\it the level}), one defines 
{\it the space of conformal blocks of level $\ell$} 
$$
CB_\lambda^\ell(z) = V_\lambda^s\cap\text{Ker}\ e(z)^{\ell - \lambda_1 + \lambda_m +1}.
$$

{\bf 2.6.} {\it Theorem.} {\it For all $z\in Z$ and $\lambda\in\CQ_m(N)$ 

(a) $p_\lambda(z)\in V^s$.

(b) If $\lambda_1 > \lambda_m$ then $e(z)p_\lambda(z) = 0$; if $\lambda_m = \lambda_1$ then $e(z)^2p_\lambda(z) = 0$. 

(c) Suppose that $\lambda_1 - \lambda_m\leq 1$ (in this case $\dim CB_\lambda^1(z) = 1$). Then 
$p_\lambda(z)$ satisfies to the system of Knizhnik - Zamolodchikov differential equations 
$$
\frac{\dpar p_\lambda(z)}{\dpar z_i} = \frac{1}{m+1}\sum_{j\neq i}\ 
\frac{\pi_{ij} - m\cdot \text{Id}}{z_i - z_j}p_\lambda(z)
$$
where $\pi\in \text{End}(\bar V\otimes \bar V),\ \pi(x\otimes y) = y\otimes x$ (note that 
$\pi = \sum_{a<b}\ e_{ab}\otimes e_{ba}$), $\pi_{ij}$ means the transposition of the $i$-th and $j$-th factors.}

\bigskip\bigskip

\newpage

\centerline{\bf \S 3. Hypergeometric solutions}

\bigskip\bigskip

In this section we recall the main construction from [SV]. 

{\bf 3.1.} {\it Master function.} Let $\fg$ be a simple complex Lie algebra of rank $r$. We fix a triangular decomposition 
$\fg = \fn_-\oplus\fh\oplus\fn_+$, the generators $e_i$ (resp. $f_i$) of $\fn_+$ 
(resp. of $\fn_-$), simple roots $\alpha_i\in\fh^*$, $i=1,\ldots,r$. 

Given a nonzero complex number $\kappa$, $N$ weights $\Lambda_j\in\fh^*,\ j=1,\ldots,N$ and a weight 
$$
\mu = \sum_{j=1}^N\ \Lambda_j - \sum_{i=1}^r\ n_i\alpha_i
\eqno{(3.1)}
$$
where all $n_i\in \BN$, we associate to these data a maultivalued {\it master function} 
$\Phi(t,z)$. It depends on two groups of variables: $z = (z_1,\ldots,z_N)$ and 
\newline $t = (t_{ia},\ 1\leq i \leq r,\ 1\leq a\leq n_i)$, and is defined by 
$$
\Phi(t,z) = \prod_{i<j}\prod_{a, b} 
(t_{ia} - t_{jb})^{(\alpha_i,\alpha_j)/\kappa}
\prod_{i}\prod_{a<b} 
(t_{ia} - t_{ib})^{(\alpha_i,\alpha_i)/\kappa}\cdot
$$
$$
\cdot \prod_{i,a}\prod_{j} 
(t_{ia} - z_j)^{-(\alpha_i,\Lambda_j)/\kappa}\cdot
\prod_{j<k} 
(z_j - z_k)^{(\Lambda_j,\Lambda_k)/\kappa}.
\eqno{(3.2)}
$$ 

{\bf 3.2.} {\it Accompanying rational functions.} Let $U$ denote the universal enveloping 
algebra  
of the free Lie algebra in generators $\tf_i,\ 1\leq i \leq r$, i.e. the free associative 
$\BC$-algebra in generators $\tf_i$. Consider its tensor power $U^{\otimes N}$. 

This algebra is $\BN^r$-graded. Namely, 
for $\bn = (n_1,\ldots,n_r)\in\BN^r$, we denote by 
$(U^{\otimes N})_\bn\subset U^{\otimes N}$ the linear subspace generated by all monomials 
$$
m = \prod \tf_j\otimes\ldots\otimes \prod \tf_k
\eqno{(*)}
$$
which contain $n_i$ times the character $\tf_i$. We denote $S_{\bn} = \prod_{i=1}^r\ S_{n_i}$.   

We associate to $m$ a rational function $\psi(m) = \psi(m;t,z)$; here $z = (z_1,\ldots,z_N)$ and $t = (t_{ia})$ is a group of variables as in 3.1.  
Note that $S_{\bn}$ acts in the evident way on the 
set $\{t_{ia}\}$; our functions $\psi(m)$ will be symmetric with respect to this action. 

First we define their "nonsymmetric" version, rational functions $\tpsi(m)$.  

By definition, $\tpsi(1\otimes\ldots\otimes 1) = 1$. We procede the definition 
by induction on the length of $m := \sum n_i$. We denote by $f_j^{(n)}\in \text{Hom}(U^{\otimes N},U^{\otimes N})$ the left multiplication by $f_j$ on the $n$-th factor. 

Let   
$m' = \tf_j^{(n)} m'$ where $\tpsi(m')$ is already defined; let $m'_n\in U$ be the 
$n$-th factor of $m'$.  
We set:  
$$
\tpsi(m) = \frac{1}{t_{jn_j} - t}\cdot \tpsi(m')
$$
where $t = z_n$ if $m'_n = 1$, $t = t_{kn_k}$ if $m'_n = \tf_k y,\ k\neq j$ and 
$t = t_{j,n_j-1}$ if $m'_n = \tf_j y$. 

Finally we set 
$$
\psi(m) = \sum_{\sigma\in S_\bn}\ \sigma\tpsi(m)
$$
where the group $S_\bn$ acts on functions $\tpsi(m)$ through variables 
$t_{ia}$. 

{\bf 3.3.} {\it Canonical element.} 
For a fixed $\bn = (n_1,\ldots,n_r)$ let 
$\{m_\alpha\}_{\alpha\in A}$ be the basis of the space $(U^{\otimes N})_\bn$ consisting of 
monomials of the form (*). To each $m_\alpha$ corresponds a rational function 
$\psi(m_\alpha)\in \BC(t,z)$; denote 
$$
\omega_\alpha := \psi(m_\alpha)\Phi(t,z)dt_{1n_1}\wedge\ldots\wedge dt_{rn_r}.
$$
This is a maultivalued differential form of degree $\ell(n):= \sum_{i=1}^r n_i$ on the 
complex affine space with corrdinates $t_{ia}, z_n$ with logarithmic singularities 
along the hyperplanes $t_{ia} = t_{jb},\ t_{ia} = z_n$. Let us denote the space 
of such forms $\Omega^{\ell(\bn)}(t,z)$ and consider an element 
$$
\tilde\delta := \sum_{\alpha\in A}\ m_\alpha\otimes \omega_\alpha \in 
(U^{\otimes N})_\bn \otimes \Omega^{\ell(\bn)}(t,z).
$$

Given $N$ weights $\Lambda_1,\ldots, \Lambda_N$ as above, let $L(\Lambda_j)$ 
denote the irreducible $\fg$-module of highest weight $\Lambda_j$, with 
the vacuum vector $1_j$. Let $\pi_j:\ U\lra L(\Lambda_j)$ be the unique epimorphism 
such that $\pi_j(1) = 1_j$ and $\pi_j(\tf_i x) = f_i\pi_j(x)$ for all $i$ and 
$x\in U$; taking their tensor product we get an epimorphism 
$$
\pi:\ U^{\otimes N} \lra L(\Lambda_1)\otimes\ldots\otimes L(\Lambda_N)
$$
which maps $(U^{\otimes N})_\bn$ onto $(L(\Lambda_1)\otimes\ldots\otimes L(\Lambda_N))_\mu$ 
where $\mu = \sum_j\ \Lambda_j - \sum_i\ n_i\alpha_i$ as in (3.1). 

We set 
$$
\delta := \pi(\tilde\delta) = \sum_{\alpha\in A}\ \pi(m_\alpha)\otimes \omega_\alpha \in 
(L(\Lambda_1)\otimes\ldots\otimes L(\Lambda_N))_\mu \otimes \Omega^{\ell(\bn)}(t,z).
$$
Finally, if $ C = \{C(z)\}$ is a family of homology cycles with coefficients 
in the local system dual to (3.2) which is horizontal with respect to Gauss-Manin connection along $z$, we can integrate $\delta$ along $C$ and get
a maultivaluedd function 
$$
\phi(z) := \int_{C(z)}\ \delta \in (L(\Lambda_1)\otimes\ldots\otimes L(\Lambda_N))_\mu.
$$
The main result of [SV] and [FSV] says that $\phi(z)$ is a section of the subbundle 
of conformal blocks (in particular, for each $\phi(z)$ is a singular vector for 
each $z$), and this section is horizontal with respect to the KZ connection, i.e. 
it satisfies to the following system of linear differential equations: 
$$
\kappa \frac{\dpar\phi}{\dpar z_j} = \sum_{i\neq j}\ \frac{\Omega_{ij}}{z_j - z_i} \phi(z),\ j = 1,\ldots, N
\eqno{(3.3)}
$$
Here we note that our simple Lie algebra $\fg$ comes equipped with a canonical 
$\fg$-invariant {\it Casimir} element $\Omega\in\fg\otimes\fg$ 
and by definition $\Omega_{ij}\in \text{End}(L(\Lambda_1)\otimes\ldots\otimes  
L(\Lambda_m))$ denotes an endomorphism acting as $\Omega$ on the product of the $i$-th and 
the $j$-th factors, and as the identity on the others. 
 
So we get a family of solutions of KZ equations numbered by the above homology cycles.

{\bf 3.4.} 
Note that given $\phi(z)$ satisfying (3.3) and any symmetric $m\times m$ 
complex matrix $(c_{ij})$, if we put 
$$
\psi(z) = \prod_{i<j} (z_i - z_j)^{c_{ij}}\phi(z)
$$
then this new function satisfies the differential equations
$$
\kappa \frac{\dpar\psi}{\dpar z_j} = \sum_{i\neq j}\ \frac{\Omega_{ij} + c_{ij}\text{Id}}{z_j - z_i} \psi(z),\ j = 1,\ldots, N
\eqno{(3.4)}
$$
The systems (3.3) and (3.4) are called {\it gauge equivalent}. 
We will use this simple remark several times. 

{\bf 3.5.} Below we will use the KZ equations for the reductive Lie algebra 
$\fgl(m)$ which look exactly as in (3.3) but now $\phi(z)\in \otimes_{i=1}^N L_i$ 
where $L_i$ are some representions or $\fgl(m)$ and $\Omega_{ij}$ is defined 
starting from 
$$
\Omega = \sum_{a, b = 1}^m\ e_{ab}\otimes e_{ba}\in \fgl(m)\otimes \fgl(m)
$$
On the tensor square of the vector representation $\bV\otimes\bV$ this 
$\Omega$ acts as $\Omega(x\otimes y) = \pi(x\otimes y):= y\otimes x$. 

On the other hand, the standard Casimir for the simple Lie algebra 
$\fsl(m)$ acts on $\bV\otimes\bV$ as $\pi - m^{-1}\cdot\text{Id}$.
 
It follows that if all $L_i = \bV$ then the KZ equations for $\fgl(m)$ and 
for $\fsl(m)$ are gauge equivalent. 

\bigskip\bigskip

\newpage

\centerline{\bf \S 4. Selberg integrals}

\bigskip\bigskip

Here we specify the previous construction to our case. 

{\bf 4.1.}  
We have $\fg = \frak{sl}(m),\ r = m - 1$, $e_i = e_{i,i+1},\ f_i = e_{i+1,i}$. 
Let us reconcile our present notation with that from 2.1.
  
We consider the vector representation $\bV = L(\omega_1)$ where $\omega_1$ is the first 
fundamental weight. It has a basis $\{y_1,\ldots,y_m\}$ where $y_1$ is a vacuum vector, 
i.e. $e_iy_1 = 0$ for all $i$, and 
$y_{j+1} = f_jy_j,\ j = 1,\ldots, m-1$. 
In other words, 
$$
y_j = f_{j-1}f_{j-2}\ldots f_1y_1.
\eqno{(4.1)}
$$
(Alternatively, we can remark that $y_j = e_{j1}y_1$ and 
$$
e_{j1} = [e_{j,j-1},[e_{j-1,j-2},\ldots [e_{32},e_{21}]\ldots ]). 
$$

\bigskip

(a) {\it The case $G/B$.}

\bigskip

{\bf 4.2.} 
Consider the weight subspace 
$V_\lambda = (\bar V^{\otimes m})_\lambda$ where $\lambda = (1,\ldots,1) \in \CP_m(m)$; 
its dimension is $m!$ and it admits a basis $\{y_\sigma\},\ \sigma\in S_m$, where 
$y_\sigma = y_{\sigma(1)}\otimes\ldots\otimes y_{\sigma(m)}$. 

The subspace of singular vectors 
$$
V^s_\lambda = \{x\in V_\lambda|\ e_ix = 0\ \text{for all}\ i\}
$$
is one-dimensional and is spanned by the vector 
$$
w = \sum_{\sigma\in S_m}\ (-1)^{|\sigma|}y_\sigma.
$$
Let us consider the following 
$V^s_\lambda$-valued Knizhnik-Zamolodchikov equation 
$$
\frac{\dpar\Psi}{\dpar z_i} = \frac{1}{\kappa}\sum_{j\neq i}\ 
\frac{\pi_{ij} - \text{Id}}{z_i - z_j}\Psi,\ 
1\leq i \leq m
\eqno{(4.2)}
$$
where $\Psi(z) = \psi(z)w\in V^s_\lambda$ and $\pi_{ij}$ is the permutation of 
$i$-th and $j$-th factor as usual (it is gauge equivalent to the standard 
$V^s_\lambda$-valued KZ equation, cf. 3.5). It has a solution 
$$
\psi(z_1,\ldots,z_m) = \prod_{1\leq i < j\leq m}\ (z_j - z_i)^{-2/\kappa}
$$

{\bf 4.3.} We see that after writing the basis vectors in the form (4.1) 
the weight space $V_\lambda$ has $m-1$ characters $f_1$, $m-2$ characters $f_2$, ..., 
$1$ character $f_{m-1}$. So in the corresponding hypergeometric integral there is 
$d := m(m-1)/2$ coordinates $t$. 

Fix on $\BC^d$ coordinates
$$
t=(t^{(1)}_{1}, \dots, t^{(1)}_{m-1}, t^{(2)}_{1}, \dots, t^{(2)}_{m-2}, \dots, t^{(m-1)}_{1}) ,
$$
the holomorphic volume form
\bea
dt &=& dt^{(m-1)} \wedge dt^{(m-2)}\wedge\dots\wedge dt^{(1)}
\\
&=&
 dt^{(m-1)}_1 \wedge dt^{(m-2)}_1\wedge dt^{(m-2)}_2 \wedge\dots\wedge dt^{(1)}_1\wedge \dots\wedge dt^{(1)}_{m-1} ,
\eea
and the master function
\bea
\Phi(t,z) &=& \prod_{i=1}^{m-1}\prod_{j=1}^{m} (t^{(1)}_i-z_j)^{-1}
\prod_{1\leq i<j\leq m-1}(t^{(1)}_j-t^{(1)}_i)^2 \times
\\
&&
\prod_{i=1}^{m-2}\prod_{j=1}^{m-1} (t^{(2)}_i-t^{(1)}_j)^{-1}
\prod_{1\leq i<j\leq m-2}(t^{(2)}_j-t^{(2)}_i)^2 \dots
\prod_{i=1}^{1}\prod_{j=1}^{2} (t^{(m-1)}_i-t^{(m-2)}_j)^{-1}.
\eea

\bigskip

{\it Actions of symmetric groups}

\bigskip

Let the group
$S_m$ act on functions of $t,z$ by permuting the variables $z_1,\dots,z_m$,
\bea
(\sigma g)(t,z_1,\dots,z_m) = g(t, z_{\sigma^{-1}(1)}, \dots, z_{\sigma^{-1}(m)}) .
\eea
Similarly let the group
$S_{m-1}$ act on functions of $t,z$ by permuting the variables
$t^{(1)}_1$, \dots, $t^{(1)}_{m-1}$ and so on.
For a function $g(t,z)$ define the symmetrizations
\bea
{\rm Sym}_z g (t,z) = \sum_{\sigma\in S_m} (\sigma h)(t,z),
\qquad
{\rm Sym}_{t^{(1)}} g (t,z) = \sum_{\sigma\in S_{m-1}} (\sigma h)(t,z),
\qquad
\text{and so on.}
\eea

\bigskip

{\it Weight functions}

\bigskip
Set
\bea
g (t,z) &=& \left[\prod_{i=1}^{m-1}(t^{(1)}_i -z_{m-i+1}) \prod_{i=1}^{m-2}(t^{(2)}_i -t^{(1)}_i)
\prod_{i=1}^{m-3}(t^{(3)}_i -t^{(2)}_i)^{-1} \dots \prod_{i=1}^1(t^{(m-1)}_i -t^{(m-2)}_i)\right]^{-1} \!\!,
\\
\omega(t,z) &=& \text{Sym}_{t^{(1)}}\text{Sym}_{t^{(2)}}\dots \text{Sym}_{t^{(m-2)}} g(t,z) .
\eea
For any $\sigma\in S_m$, define
\bea
\omega_\sigma(t,z) = (\sigma\omega)(t,z) .
\eea

\bigskip

\noindent
{\it Example.} For $m=2$, $\sigma = id \in S_2$ and $\sigma'=(12)\in S_2$ we have
\bea
\omega_\sigma = (t^{(1)}_1-z_2)^{-1},\qquad
\omega_{\sigma'} = (t^{(1)}_1-z_1)^{-1} .
\eea
For  $m=3$, $\sigma = id \in S_3$ and $\sigma'=(13)\in S_2$ we have
\bea
\omega_\sigma &=& [(t^{(1)}_1-z_3)(t^{(1)}_2-z_2)(t^{(2)}_1-t^{(1)}_1)]^{-1}
+ [(t^{(1)}_2-z_3)(t^{(1)}_1-z_2)(t^{(2)}_1-t^{(1)}_2)]^{-1} ,
\\
\omega_{\sigma'} &=& [(t^{(1)}_1-z_1)(t^{(1)}_2-z_2)(t^{(2)}_1-t^{(1)}_1)]^{-1}
+ [(t^{(1)}_2-z_1)(t^{(1)}_1-z_2)(t^{(2)}_1-t^{(1)}_2)]^{-1} .
\eea
Define
\bea
\omega(t,z) = \sum_{\sigma\in S_m} \omega_\sigma(t,z) y_\sigma .
\eea
This is a $V_\lambda$-valued function of $t,z$.

\bigskip

{\bf 4.4.} {\it Integrals.} Consider the $V_\lambda$-valued differential $d$-form
\bea
\Phi(t,z)^{1/\kappa} \omega(t,z) dt .
\eea
Let $\delta(z)$ be a flat section of the homological bundle associated with this differential form, see [SV, V3]. Then by [SV] 
the $V_\lambda$-valued function
\bea
I(z) = \int_{\delta(z)} \Phi(t,z)^{1/\kappa} \omega(t,z) dt
\eea
takes values in the space of singular vectors $V^s_\lambda$ and is a solution of the KZ equations. 

\bigskip

{\bf 4.5.} {\it Gelfand-Zetlin cycle.}


For real $z=(z_1,\dots,z_m)$ with $z_1<z_2<\dots <z_m$ define a $d$-dimensional cell
\bea
\gamma_m = \gamma_m(t;z) = \gamma_m(t^{(1)},\dots,t^{(m-1)},z)
\eea
in $\BC^d$ by the conditions
\bea
&t^{(m-2)}_1 < t^{(m-1)}_1 < t^{(m-2)}_2 ,&
\\
&t^{(m-3)}_1 < t^{(m-2)}_1 < t^{(m-3)}_2 < t^{(m-2)}_2<t^{(m-3)}_3 , &
\\
&\dots &
\\
&t^{(1)}_1 < t^{(2)}_1 < t^{(1)}_2 < \dots <
t^{(1)}_{m-2} < t^{(2)}_{m-2} < t^{(1)}_{m-1} , &
\\
&z_1 < t^{(1)}_1 < z_2 < \dots <
z_{m-1} < t^{(1)}_{m-1} < z_{1} . &
\eea
We denote by $\gamma_m^{m-1}(t^{(m-2)})$ the set of all points $t^{(m-1)}$  satisfying the conditions in the first line of these inequalities.
We denote by $\gamma_m^{m-2}(t^{(m-3)})$ the set of all points $t^{(m-2)}$ satisfying the conditions  in the second line of these inequalities and so on until
we denote by $\gamma_m^{1}(z)$ the set of all points  $t^{(1)}$ satisfying the conditions in the last line of these inequalities.

\bigskip

The Gelfand - Zetlin cell $ \gamma_m(t^{(1)},\dots,t^{(m-1)};z)$ has an important {\it factorization property}: 

$\gamma_m(t^{(1)},\dots,t^{(m-1)};z)$ consists of points
$(t^{(1)},\dots,t^{(m-1)})$ such that
 $t^{(1)}$ lies in
$\gamma_m^{1}(z)$ and $(t^{(2)},\dots,t^{(m-1)})$ lies in
$\gamma_{m-1}(t^{(2)},\dots,t^{(m-1)};t^{(1)})$. 

\bigskip

{\bf 4.6.} Consider the iterated integral over $\gamma_m(t;z)$,
\bea
I_\kappa(z) =\int_{\gamma^1_m(z)} dt^{(1)}
\int_{\gamma^2_m(t^{(1)})} dt^{(2)}
\dots
\int_{\gamma^{m-1}_m(t^{(m-2)})} dt^{(m-1)}
 \Phi(t,z)^{1/\kappa} \omega(t,z) . 
\eea
The function $\Phi^{1/\kappa}$ is multivalued.
In order to define the integral (apart from its possible divergence)
we need to choose a section over $\gamma_m(t,z)$ of the local system associated with the function $\Phi^{1/\kappa}$.
We choose the section
\bea
&&
\prod_{i=1}^{m-1}\prod_{j=1}^{m} |t^{(1)}_i-z_j|^{-1/\kappa}
\prod_{1\leq i<j\leq m-1}|t^{(1)}_j-t^{(1)}_i|^{2/\kappa} \times
\\
&&
\prod_{i=1}^{m-2}\prod_{j=1}^{m-1} |t^{(2)}_i-t^{(1)}_j|^{-1/\kappa}
\prod_{1\leq i<j\leq m-2}|t^{(2)}_j-t^{(2)}_i|^{2/\kappa} \dots
\prod_{i=1}^{1}\prod_{j=1}^{2} |t^{(m-1)}_i-t^{(m-2)}_j|^{-1/\kappa}.
\eea


{\bf 4.7.} {\it Theorem.}  
{\it Let $z_1<z_2<\dots<z_m$ and $1/\kappa<0$. Then the integral $I_\kappa(z)$ is convergent and equals
$C_m(\kappa)\psi(z) w$, where
\bea
C_m(\kappa) = (-1)^{a_m}\frac {\Gamma(1-1/\kappa)^{m(m+1)/2}}
{(-1/\kappa)^{m-1} \prod_{i=1}^m \Gamma(1-i/\kappa)} ,
\qquad
a_m=m-1+\sum_{j=4}^m {m-2\choose 2} .
\eea
Moreover, the integral has a well-defined analytic continuation with respect to $1/\kappa$ to the region where the real part of $1/\kappa$ is less than
$ 1/m$.}


{\bf 4.8.} {\it Proof} is by induction on  $m$. For $m=2$ we have
\bea
\Phi = |t^{(1)}_1-z_1|^{-1}|t^{(1)}_1-z_2|^{-1},
\eea
\bea
I_\kappa(z_1,z_2)
&=&
\int_{z_1}^{z_2} dt^{(1)}_1\Phi^{1/\kappa}
(\frac{v_{id}}{t^{(1)}_1-z_2} + \frac{v_{(21)}}{t^{(1)}_1-z_1})
\\
&=&
(z_1-z_2)^{-2/\kappa} \int_0^1 dt\ t^{-1/\kappa}(1-t)^{-1/\kappa}
(\frac{v_{id}}{t-1} + \frac{v_{(21)}}{t})
\\
&=&
(z_1-z_2)^{-2/\kappa} \frac{\Gamma(1-1/\kappa)
\Gamma(-1/\kappa)}{\Gamma(1-2/\kappa)}
(-v_{id}+ v_{(21)})
\eea
and the first statement of the  theorem is proved for $m=2$. To show the required analytic continuation we replace the integral over the interval by
the corresponding Pochhammer double loop.

Let $m=3$. The integral is three-dimensional. The first integration over $t^{(2)}_1$ from $t^{(1)}_1$ to
$t^{(1)}_2$ is exactly the calculation of the integral for $m=2$ in which $z_1,z_2,t^{(1)}_1$ are replaced with $t^{(1)}_1,
t^{(1)}_2,t^{(2)}_1$, respectively. Using the result for $m=2$,
we see that after the first integration in the remaining double integral over
$t^{(1)}_1$, $t^{(1)}_2$ the factor $(t^{(1)}_2-t^{(1)}_1)^{2/\kappa}$ in the master function
is canceled with the factor $(t^{(1)}_2-t^{(1)}_1)^{2/\kappa}$ obtained after the first integration. Therefore, in the remaining double integral
the variables $t^{(1)}_1$, $t^{(1)}_2$ become decoupled. More precisely, we have
\bea
I_\kappa(z_1,z_2,z_3) = \sum_{\sigma\in S_3} I_{\kappa,\sigma}(z_1,z_2,z_3) y_\sigma
\eea
where $I_{\kappa,\sigma}$ is the determinant of the $2\times 2$-matrix whose rows are
\bea
&&
\int_{z_1}^{z_2} \tilde\Phi (t^{(1)}_1,z)^{1/\kappa} (t^{(1)}_1-z_a)^{-1} dt^{(1)}_1
\qquad
\int_{z_1}^{z_2} \tilde\Phi (t^{(1)}_1,z)^{1/\kappa} (t^{(1)}_1-z_b)^{-1} dt^{(1)}_1 \ ,
\\
&&
\int_{z_2}^{z_3} \tilde\Phi(t^{(1)}_2,z)^{1/\kappa} (t^{(1)}_2-z_a)^{-1} dt^{(1)}_2
\qquad
\int_{z_2}^{z_3} \tilde\Phi (t^{(1)}_2,z)^{1/\kappa} (t^{(1)}_2-z_b)^{-1} dt^{(1)}_2
\eea
with $a=\sigma^{-1}(3),\ b=\sigma^{-1}(2)$, $\tilde \Phi(s,z) = \prod_{i=1}^3(s-z_i)^{-1}$.
By [V1] this determinant equals
\bea
-(-1)^\sigma\frac {\Gamma(1-1/\kappa)\Gamma(1-1/\kappa)\Gamma(-1/\kappa)}
{\Gamma(1-3/\kappa)} ,
\eea
see also Section 3.3 in [V2].
Together with the statement for $m=2$, this formula implies the first statement of the theorem for $m=3$.
To show the required analytic continuation we again replace integrals over the intervals by
the corresponding Pochhammer double loops.

For arbitrary $m$ we use the induction hypothesis and reduce the coefficients of the basis vectors to
$(m-1)\times (m-1)$-determinants of one-dimensional integrals. Those determinants were calculated in
[V1], cf.  [V2]. As a result we get the first statement of the theorem for arbitrary $m$. The second statement is proved by using
the Pochhammer double loops. $\square$

\bigskip

(b) {\it Selberg integrals associated with conformal blocks at level 1}

\bigskip

{\bf 4.9.} {\it Subbundle of conformal blocks of level 1.} Now for arbitrary $N$ 
consider the $N$-th tensor power of $V = \bV^{\otimes N}$ where $\bV$ is the vector 
representation of $\frak{sl}(m)$ as before. Denote by
$V_\lambda$ its weight subspace of weight
\bea
\lambda =(\la_1,\dots,\la_m) =
(a+1,\dots,a+1,a\dots,a)= (1,\dots,1,0,\dots,0)+(a,\dots,a)
\eea
where $a$ is some nonnegative integer,
the vector $(1,\dots,1,0,\dots,0)$ has $m'$ ones with $0\leq m' < m$.

For $z=(z_1,\dots,z_N)$ with distinct coordinates, denote by $CB^1(z)\subset V_\lambda$ the corresponding
one-dimensional subspace of conformal blocks of level $1$; it has rank $1$ and 
admits as a generating section
\bea
\label{can elt}
\Psi(z) = p_\lambda(z) = \sum_{w\in S(\lambda)}\ \frac{y^w}{e_w}, 
\eea
cf.  2.4. That section is a solution of the KZ differential equations
\bea
\label{KZ1}
\frac{\der \Psi}{\der z_i} \ = \ \frac 1{m+1} \,\sum_{j\neq i} \frac{\pi_{ij} - m\cdot{\Id}}{z_i-z_j}
\, \Psi, \qquad i=1,\dots,N.
\eea
Notice that the coefficient of $\Id$ in these equations is different from the coefficient of $\Id$ in (4.2).

{\bf 4.10.} {\it Master function.}
Denote  $\mu_i = \la_{i+1}+\dots+\la_m$ for $i=0,\dots,m-1$, and
\bea
d_N = \la_2 + 2\la_3+\dots+(m-1)\la_m = \mu_1+\dots +\mu_{m-1} =  a\frac {m(m-1)}2 + \frac {m'(m'-1)}2 .
\eea
Fix on $\C^{d_N}$ coordinates $t=(t^{(1)}, \dots, t^{(m-1)})$, where
\bea
t^{(i)} = (t^{(i)}_1,\dots,t^{(i)}_{\mu_{i}}) ,
\qquad
i=1,\dots,m-1.
\eea
Fix
the holomorphic volume form on $\C^{d_N}$,
\bea
dt &=& dt^{(m-1)} \wedge dt^{(m-2)}\wedge\dots\wedge dt^{(1)}
\\
&=&
 dt^{(m-1)}_1 \wedge \dots \wedge dt^{(m-1)}_{\mu_{m-1}} \wedge
 \dots \wedge dt^{(1)}_1\wedge \dots\wedge dt^{(1)}_{\mu_1} ,
\eea
and the master function
\bea
\Phi(t,z) &=&
\prod_{1\leq i<j\leq N}(z_j-z_i)^{1-m}
\prod_{i=1}^{\mu_1}\prod_{j=1}^{N} (t^{(1)}_i-z_j)^{-1}
\prod_{1\leq i<j\leq \mu_1}(t^{(1)}_j-t^{(1)}_i)^2 \times
\\
&&
\prod_{i=1}^{\mu_2}\prod_{j=1}^{\mu_1} (t^{(2)}_i-t^{(1)}_j)^{-1}
\prod_{1\leq i<j\leq \mu_2}(t^{(2)}_j-t^{(2)}_i)^2 \dots \times
\\
&&
\prod_{i=1}^{\mu_{m-1}}\prod_{j=1}^{\mu_{m-2}} (t^{(m-1)}_i-t^{(m-2)}_j)^{-1}
\prod_{1\leq i<j\leq \mu_{m-1}} (t^{(m-1)}_j-t^{(m-1)}_i)^2 \ .
\eea

{\bf 4.11.} {\it Actions of symmetric groups.}
Let the group
$S_N$ act on functions of $t,z$ by permuting the variables $z_1,\dots,z_N$,
\bea
(\sigma g)(t,z_1,\dots,z_N) = g(t, z_{\sigma^{-1}(1)}, \dots, z_{\sigma^{-1}(N)}) .
\eea
Similarly let the group
$S_{\mu_1}$ act on functions of $t,z$ by permuting the variables
$t^{(1)}_1$, \dots, $t^{(1)}_{\mu_1}$ and so on.
For a function $g(t,z)$ define the symmetrizations
\bea
{\rm Sym}_z g (t,z) = \sum_{\sigma\in S_N} (\sigma h)(t,z),
\qquad
{\rm Sym}_{t^{(1)}} g (t,z) = \sum_{\sigma\in S_{\mu_1}} (\sigma h)(t,z),
\qquad
\text{and so on.}
\eea

{\bf 4.12.} {\it Weight functions.}
Let $v_I$ be a basis vector of $V_\lambda$. We have $I=(I_1,\dots,I_m)$ where $I_1,\dots,I_m$ form a partition of
the set $\{1,\dots,N\}$ with $|I_i|=\la_i$, $i=1,\dots,m$.

For every $i=0,\dots,m-1$, fix a bijection
\bea
\nu_j : \{1,\dots,\mu_i\} \to I_{i+1}\cup\dots\cup I_m
\eea
such that the first $\la_m$ elements of $\{1,\dots,\mu_i\}$ are mapped to $I_m$, the next
$\la_{m-1}$ elements of $\{1,\dots,\mu_i\}$ are mapped to $I_{m-1}$ and so on until the last
$\la_{i+1}$ elements of $\{1,\dots,\mu_i\}$ are mapped to $I_{i+1}$.

Denote
\bea
g_{I,\nu}(t,z)
 &=& \prod_{i=1}^{\mu_1}(t^{(1)}_{\nu_1(i)} -z_{\nu_0(i)})^{-1} \prod_{i=1}^{\mu_2}(t^{(2)}_{\nu_2(i)} -t^{(1)}_{\nu_1(i)})^{-1}
 \dots
 \prod_{i=1}^{\mu_{m-1}}(t^{(m-1)}_{\nu_{m-1}(i)} -t^{(m-2)}_{\nu_{m-2}(i)})^{-1} ,
\\
\omega_{I}(t,z) &=& \text{Sym}_{t^{(1)}}\text{Sym}_{t^{(2)}}\dots \text{Sym}_{t^{(m-2)}} g_{I,\nu}(t,z) ,
\\
\omega(t,z) &=& \sum_{I} \omega_I(t,z) v_I .
\eea
This is a $V_\lambda$-valued function of $t,z$.

{\bf 4.13.} {\it Integrals.} Consider the $V_\lambda$-valued differential $d_N$-form
\bea
\Phi(t,z)^{1/\kappa} \omega(t,z) dt .
\eea
Let $\delta(z)$ be a flat section of the homological bundle associated with this differential form, see [SV], [V3]. Then by [SV], [FSV]
the $V_\lambda$-valued function
\bea
I(z) = \int_{\delta(z)} \Phi(t,z)^{1/\kappa} \omega(t,z) dt
\eea
is a solution of the KZ equations
\bea
\label{KZ2}
\frac{\der I}{\der z_i} \ = \ \frac 1{\kappa} \,\sum_{j\neq i} \frac{\pi_{ij} - m{\Id}}{z_i-z_j}
\, I, \qquad i=1,\dots,N,
\eea
moreover if $\kappa = m+1$, then $I(z)\in CB^1(z)$.

{\bf 4.14.} {\it The cycle.}
\label{Sec Cycle}
For real $z=(z_1,\dots,z_N)$ with $z_1<z_2<\dots <z_N$ we define a $d_N$-dimensional cell
$
\gamma = \gamma(t;z) = \gamma(t^{(1)},\dots,t^{(m-1)},z)
$
in $\C^{d_N}$ as follows.

We split numbers $z_1,\dots,z_N$ into $a+1$ groups
\bea
z^{(j)}&=&(z_{m(j-1)+1},\dots, z_{m(j-1)+m}), \qquad
 j=1,\dots,a,
 \\
 z^{(a+1)}&=&(z_{ma+1}, \dots, z_{ma+m'}=z_N) .
 \eea
  We split
variables $t^{(i)}$ into $a+1$ groups
\bea
t^{(i,j)} &=& (t^{(i)}_{(m-i)(j-1)+1},\dots,t^{(i)}_{(m-i)(j-1)+(j-1)}),
\qquad
 j=1,\dots,a,
\\
t^{(i,a+1)} &=& (t^{(i)}_{(m-i)(j-1)+1},\dots,t^{(i)}_{\mu_i}) .
\eea
Note that the last group $t^{(i,a+1)}$ is empty for $i\geq m'$.
We define
\bea
\gamma(t;z) &=& \gamma_m(t^{(1,1)},\dots,t^{(m-1,1)};z^{(1)})\times
 \gamma_m(t^{(1,2)},\dots,t^{(m-1,2)};z^{(2)})\times \dots
 \\
 &&
 \dots\times
  \gamma_m(t^{(1,a)},\dots,t^{(m-1,a)};z^{(a)})\times
   \gamma_{m'}(t^{(1,a+1)},\dots,t^{(m'-1,a+1)};z^{(a+1)})
   \eea
where the cells in the right hand side are introduced in Section 4.5.

{\bf 4.15.} 
Consider the following iterated integral over the cell $\gamma(t;z)$:  
\bea 
\label{INTEGRAL general}
I_\kappa(z) =\int dt^{(1)}
\int dt^{(2)}
\dots
\int dt^{(m-1)}
 \Phi(t,z)^{1/\kappa} \omega(t,z),   
\eea
cf. 4.6. 
The function $\Phi^{1/\kappa}$ is multivalued.
In order to define the integral (apart from its possible divergence)
 we need to choose a section over the cell $\gamma(t;z)$ of the local system associated with the function $\Phi^{1/\kappa}$.
We choose the section
\bea
&&
\prod_{1\leq i<j\leq N}|z_j-z_i|^{(1-m)/\kappa}
\prod_{i=1}^{\mu_1}\prod_{j=1}^{N} |t^{(1)}_i-z_j|^{-1/\kappa}
\prod_{1\leq i<j\leq \mu_1}|t^{(1)}_j-t^{(1)}_i|^{2/\kappa} \times
\\
&&
\prod_{i=1}^{\mu_2}\prod_{j=1}^{\mu_1} |t^{(2)}_i-t^{(1)}_j|^{-1/\kappa}
\prod_{1\leq i<j\leq \mu_2}|t^{(2)}_j-t^{(2)}_i|^{2/\kappa} \dots \times
\\
&&
\prod_{i=1}^{\mu_{m-1}}\prod_{j=1}^{\mu_{m-2}} |t^{(m-1)}_i-t^{(m-2)}_j|^{-1/\kappa}
\prod_{1\leq i<j\leq \mu_{m-1}} |t^{(m-1)}_j-t^{(m-1)}_i|^{2/\kappa} \ .
\eea


{\bf 4.16.} {\it Theorem.} 
{\it For $z_1<z_2<\dots<z_N$  and $1/\kappa<0$ the integral $I_\kappa(z)$ in 4.15 is convergent
and is a solution to the KZ equations from 4.9. 

The integral $I_\kappa(z)$ has a well-defined analytic continuation
to $\kappa=m+1$.
We have
\bea
I_{m+1}(z) = C p_\lambda(z), \qquad \text{with}\
C \,=\, (C_m(m+1))^aC_{m'}(m+1),
\eea
where the constants $C_m(m+1)$, $C_{m'}(m+1)$ are defined in Theorem 4.7.}

{\bf 4.17.} {\it Proof.} The first two statements of the theorem follow from \S 3.

For $\kappa=m+1$, both functions
$I_{m+1}(z)$ and $p_\lambda(z)$ are solutions of the KZ equations with values in the one-dimensional bundle of conformal blocks, hence, they are proportional.
The coefficient of the proportianality is calculated in the limit $z^{(i)}\to y_i$, $i=1,\dots,a+1$, where
$y_1<\dots<y_{a+1}$ are some fixed numbers. Comparing the asymptotics of both functions we get the formula for $C$, cf. Theorem 4.8 in [RV]. $\square$

\bigskip\bigskip



\centerline{\bf \S 5. Action of positive currents} 

\bigskip\bigskip

{\bf 5.1.} Let us return to the setup of \S 2. For each $\lambda\in\CP_m(N)$ we have 
defined an isomorphism of (free) $R'$-modules 
$$
H^*_T(X_\lambda)_{R'}\overset{\sim}\lra (\bV^{\otimes N})_{\lambda;R'}.
\eqno{(5.1)}
$$
Let us denote by $X_{m,N}$ the variety of all flags 
$$
0\subset L_1\subset\ldots\subset L_m = \BC^N
\eqno{(5.2)}
$$ 
of length $m$ in $\BC^N$, so it is  the disjoint union 
$$
X_{m,N} = \coprod_{\lambda\in\CP_m(N)}\ X_\lambda.
$$
Summing up (5.1) over all $\lambda\in\CP_m(N)$ we get an isomorphism of $R'$-modules 
$$
H^*_T(X_{m,N})_{R'}\overset{\sim}\lra (\bV^{\otimes N})_{R'} = \bV^{\otimes N}\otimes_\BC R'.
\eqno{(5.2)}
$$
The Lie algebra $\fgl(m)$ acts on $(\bV^{\otimes N})_{R'}$ through its action on 
$\bV^{\otimes N}$. Due to the extension of scalars one can extend this action to an action 
of the Lie algebra of positive currents $\fgl(m)[t]$. Namely, for $x\in \fgl(m)$ 
the action of $xt^j$ on $\bV^{\otimes N}$ is defined by the operator
$$
xt^j = \sum_{i=1}^N\ x^{(i)}z_i^j
$$
(as is usual in Conformal Field Theory, one should imagine the $i$-th tensor factor 
of $\bV^{\otimes N}$ as sitting at a point $z_i$ of the Riemann sphere).  
 
In this section we shall define geometrically an action of $\fgl(m)[t]$ on the equivariant 
cohomology $H^*_T(X_{m,N})$ in such a way that after the extension of scalars to 
$R'$ the isomorphism (5.1) will be compatible with this action. 

{\bf 5.2.} Given $\lambda = (\lambda_1,\ldots,\lambda_m)\in\CP_m(N)$ and $1\leq a < m$, set 
$$
e_{a,a+1}\lambda = (\lambda_1,\ldots,\lambda_{a-1},\lambda_{a}+1,\lambda_{a+1}-1,
\lambda_{a+2},\ldots,\lambda_m),
$$
this is defined if $\lambda_{a+1}>0$, and 
$$
e_{a+1,a}\lambda = (\lambda_1,\ldots,\lambda_{a-1},\lambda_{a}-1,\lambda_{a+1}+1,
\lambda_{a+2},\ldots,\lambda_m),
$$
this is defined if $\lambda_{a}>0$. Recall that $X_\lambda$ parametrizes flags 
(5.2) with $\mu_i:= \dim L_i$ such that $\lambda_i = \mu_i - \mu_{i-1}$. 

Define 
$$
\mu'(\lambda,a) = (\lambda_1,\ldots,\lambda_{a-1},\lambda_{a},1,\lambda_{a+1}-1,
\lambda_{a+2},\ldots,\lambda_m)\in\CP_{m+1}(N)
$$
and 
$$
\mu''(\lambda,a) = (\lambda_1,\ldots,\lambda_{a-1},\lambda_{a}-1,1,\lambda_{a+1},
\lambda_{a+2},\ldots,\lambda_m)\in\CP_{m+1}(N)
$$
Consider the variety $X'_{\lambda,a} := X_{\mu'(\lambda,a)}$. We have obvious projections 
$$
X_\lambda\overset{\pi'_1}{\lla} X'_{\lambda,a}\overset{\pi'_2}\lra X_{e_{a,a+1}\lambda}.
$$
Let $S'$ (resp. $Q'$) denote the rank $1$ (resp. rank $\lambda_{a+1}-1$) vector bundle over 
$X'_{\lambda,a}$ whose fiber over a flag $L_1\subset\ldots\subset L_{m+1}=\BC^N$ is 
$L_{a+1}/L_a$ (resp. $L_{a+2}/L_{a+1}$). 

We define the map 
$$
\rho(e_{a,a+1}t^j):\ H^*_T(X_\lambda) \lra H^*_T(X_{e_{a,a+1}\lambda})
$$
by
$$
\rho(e_{a,a+1}t^j)(x) = \pi'_{2*}(\pi_1^{\prime *}(x)\cdot e(Hom(S',Q'))\cdot e(S^{\prime\otimes j}))
\eqno{(5.3)'}
$$
where $e(L)$ denotes the Euler (top Chern) class of a vector bundle $L$.       

Similarly, 
consider the variety $X''_{\lambda,a} := X_{\mu''(\lambda,a)}$. We have obvious projections 
$$
X_\lambda\overset{\pi''_1}{\lla} X''_{\lambda,a}\overset{\pi''_2}\lra X_{e_{a+1,a}\lambda}.
$$
Let $S''$ (resp. $Q''$) denote the rank $\lambda_{a}-1$ (resp. rank $1$) 
vector bundle over 
$X''_{\lambda,a}$ whose fiber over a flag $L_1\subset\ldots\subset L_{m+1}=\BC^N$ is 
$L_{a}/L_{a-1}$ (resp. $L_{a+1}/L_{a}$). 

We define the map 
$$
\rho(e_{a+1,a}t^j):\ H^*_T(X_\lambda) \lra H^*_T(X_{e_{a+1,a}\lambda})
$$
by
$$
\rho(e_{a+1,a}t^j)(x) = \pi''_{2*}(\pi_1^{\prime\prime*}(x)\cdot e(Hom(S'',Q''))\cdot e(Q^{\prime\prime\otimes j})).
\eqno{(5.3)''}
$$

Note that the maps $(5.3)'$ and $(5.3)''$ are $R = H^*_T(pt)$-equivariant, due to 
the projection formula, so they may be localized to $R'$. 

{\bf 5.3.} {\it Theorem.} {\it The maps $(5.3)'$ and $(5.3)''$ define an action of the 
Lie algebra $\fgl(m)[t]$ on $H^*_T(X_{m,N})$ such that (after extension of scalars to 
$R'$) the isomorphism (5.1) is $\fgl(m)[t]$-equivariant.}  

To prove the theorem, one remarks first that we can do the above extension of scalars. 
After that, one checks that the action of the operators $\rho(e_{a,a+1}t^j)$ and 
$\rho(e_{a+1,a}t^j)$ transferred to $\bV^{\otimes N}_{R'}$ via the isomorphism (5.1) 
coincides with the action described in 5.1. 
The details will appear elsewhere. 

{\bf 5.4.} {\it Remark.} Let us consider the cotangent bundle $T^*X_{m,N}$; it may be 
realized as  the variety of pairs $\{(L_1\subset\ldots L_m)\in X_{m,N}, A\in \text{End}(\BC^N),\ 
A(L_i)\subset L_{i-1},\ i = 2, \ldots, m\}$. This variety is of course $GL(N)$-equivariantly homotopically equivalent to $X_{m,N}$. However, it admits one more 
symmetry --- an action of $\BC^*$ by dilations along the fibers. 

The work of Ginzburg, Nakajima, Vasserot, Varagnolo, ... (cf. [CG] [N, \S 7], [V] and 
references therein) shows that the equivariant cohomology 
$H^*_{\BC^*\times GL(N)}(T^*X_{m,N})$ admits an action of the Yangian 
of the loop algebra $Y(\fgl(m)[t,t^{-1}])$. 
If we forget the action of $\BC^*$, we get the cohomology 
$H^*_{GL(N)}(T^*X_{m,N}) = H^*_{GL(N)}(X_{m,N})$ which is a quotient of 
$H^*_{\BC^*\times GL(N)}(T^*X_{m,N})$ and the above action of the Yangian 
should factor through a quotient isomorphic to $U(\fgl(m)[t])$ (this was 
explained to us by Misha Finkelberg). 

The spaces $H^*_{GL(N)}(X_{m,N})$ and $H^*_{T}(X_{m,N})$ are different but closely 
related. Namely, $T\subset GL(N)$ is a maximal torus, and 
$$
H^*_{T}(X_{m,N}) = H^*_{GL(N)}(X_{m,N})\otimes_{H^*_{GL(N)}(pt)}H^*_{T}(pt). 
$$
One should expect that the Ginzburg-Vasserot-Varagnolo action induces the 
action defined in the previous sections, however we did not verify this. 

\bigskip\bigskip



\centerline{\bf References}

\bigskip\bigskip

[AB] M.F.Atiyah, R.Bott, The moment map and equivariant cohomology, {\it Topology} {\bf 23} (1984), 1 - 28. 

[Br] A.Braverman, Instanton counting via affine Lie algebras I: 
Equivariant $J$-functions of (affine) flag manifolds and Whittaker vectors, 
{\it CRM Proc. Lecture Notes} {\bf 38}, AMS, Providence, RI (2004), 
113 - 132. 

[CG] N.Chriss, V.Ginzburg, Representation theory and complex geometry, Birkha\"user, 1997. 

[FSV] B.Feigin, V.Schechtman, A.Varchenko, On algebraic equations satisfied by hypergeometric correlators in WZW models, I: {\it Comm. Math. Phys.} 
{\bf 163} (1994), 173 - 184; II: {\it ibid.} {\bf 170} (1995), 219 - 247.  


[G1] A.Givental, Equivariant Gromov-Witten invariants, {\it IMRN} {\bf 13} 
(1996), 613 - 663.   

[G2] A.Givental, Stationary phase integrals, quantum Toda lattices, flag manifolds and the Mirror conjecture, {\it Topics in singularity theory}, 
{\it AMS Transl. Ser. 2} {\bf 180} (1997), 103 - 115. 

[N] H.Nakajima, Instantons on ALE spaces, quiver varieties, and Kac-Moody Lie algebras, 
{\it Duke Math. J.}, {\bf 76} (1994), 365 - 415. 

[RV] R.Rim\'anyi, A.Varchenko, Conformal blocks in the tensor product of vector 
representations and localization formulas, arXiv:0911.3253.  

[S] V.Schechtman, On hypergeometric functions connected with quantum cohomology 
of flag spaces, {\it Comm. Math. Phys.} {\bf 208} (1999), 355 - 379.

[SV] V.Schechtman, A.Varchenko, Arrangements of hyperplanes and Lie algebra homology, 
{\it Inv. Math.} {\bf 106} (1991), 139 - 194. 

[V] M.Varagnolo, Quiver varieties and Yangians, {\it Letters Math. Phys.} 
{\bf 53} (2000), 273 - 283.  

[V1] A.Varchenko, Euler Beta function, Vandermonde detreminant,
Legendre equation and
critical values of linear functions on a hyperplane arrangement, I, II,
Izv. Acad. Sci. USSR, Ser. Mat., {\bf 53}, no. 6 (1989), 1206 - 1234;
{\bf 54}, no. 1 (1990),

[V2] A.Varchenko, Special functions, KZ type equations, and
representation theory, AMS 2003.

[V3] A.Varchenko, Multidimensional hypergeometric functions and representation
theory of Lie algebras ans quantum groups, World Scientific, 1995.  

[W] A.Weil, Numbers of solutions of equations in finite fields, {\it Bull. AMS} {\bf 55} 
(1949), 497 - 508.

\end{document}